\documentclass[12pt,twoside]{article}
\usepackage{psfig}
\usepackage[reqno]{amsmath}
\usepackage{amssymb}
\usepackage{heronb}

\begin{document}
\title{Quantum observables associated with Einstein localisation}

\authors{Marc-Thierry Jaekel}{La\-bo\-ra\-toi\-re de Phy\-si\-que
Th\'eo\-ri\-que
\thanks{La\-bo\-ra\-toi\-re du CNRS as\-so\-ci\'e \`a
l'Eco\-le Nor\-ma\-le Su\-p\'e\-rieu\-re et \`a
l'Uni\-ver\-si\-t\'e Pa\-ris Sud},
ENS, 24 rue Lhomond, F75231 Paris Cedex 05 France}
\authors{Serge Reynaud}{Laboratoire Kastler Brossel
\thanks{La\-bo\-ra\-toi\-re de l'Eco\-le Nor\-ma\-le Su\-p\'e\-rieu\-re
et de l'Uni\-ver\-si\-t\'e Pier\-re et Ma\-rie Cu\-rie as\-so\-ci\'e au CNRS},
UPMC, case 74, 4 place Jussieu, F75252 Paris Cedex 05 France}
\maketitle

\section{Introduction}

Time and space are basic elements of our physical comprehension of the
world. They are also among the physical quantities which are measured with
the greatest accuracy. Yet their precise status raises questions in
physical theory and different notions of time and space are used
depending on the problem which is considered.

This idea is clearly emphasised by the following quotations from Newton's
{\it Principia} \cite{Newton}:

\begin{quote}
``Absolute, true, and mathematical time, of itself, and from its own nature,
flows equably without relation to anything external, and by another name is
called duration''

``Relative, apparent and common time, is some sensible and external measure
of duration by the means of motion, which is commonly used instead of true
time''

``Absolute space, in its own nature, without relation to anything external,
remains always similar and immovable''

``Relative space is some movable dimension or measure of the absolute
spaces; which our senses determine by its position to bodies; and which is
commonly taken for immovable space''
\end{quote}

These quotations show that theoretical physics has been built on at least
two different notions of time and space from its very beginning. On one
hand, time and space are defined as the mathematical parameters used to
write down the equations of motion of theoretical physics.
On the other hand, time and space are physical quantities
which are obtained through measurements.

When he introduced relativistic conceptions of space-time, Einstein
emphasised that true physical notions were that of observables.
A physical time has to be associated with an event such as the tick
of a clock or the click of a detector. He then demonstrated that
time and space observables are relativistic observables.
In particular, time and space observables are mixed under Lorentz frame
transformations so that the notion of simultaneity is no longer
absolute \cite{Einstein05}.

Remote clocks have to be synchronised through the transfer of time
references. A particularly important concept introduced by Einstein is that
of clock synchronisation through the transfer of light pulses.
This synchronisation procedure and the related localisation procedure which
consists in the exchange of several time references between different observers
\cite{IEEE91} are nowadays used for practical applications such as
the Global Positioning System \cite{GPS91} or the definition of reference
systems \cite{PetitW94}.

The previous arguments refer to classical theory of relativity but it should
be obvious that time and space observables certainly belong to the quantum
domain. The modern metrological definition of time and space has its roots
in atomic physics and is hence based on quantum theory. The time delivered
by an atomic clock is nothing but the phase of a quantum oscillator.
Electromagnetic signals used in synchronisation or localisation procedures
are quantum fields. As a consequence, any practical realisation of time has
to meet quantum limitations at some level of accuracy \cite{SaleckerW58}.

This discussion revives the basic idea contained in Newton's quotations
reproduced above, now in the context of physical theories of 20$^{{\rm th}}$
century. Physicists deal with two different notions of time and
space. The equations of motion of classical physics, but also those of
quantum field theory and of general relativity, are written in terms of
coordinate parameters, {\it i.e.} classical numbers which map space-time. A
basic assumption of general relativity is that this mapping is arbitrary.
These classical coordinate parameters necessarily differ from time and space
observables involved in any physical measurements. Space-time observables
are relativistic observables which are mixed under frame
transformations as well as quantum observables which have
quantum fluctuations and cannot be represented as classical numbers.

In this context arises the particularly acute problem of `quantum time'.
Since it is often argued that standard quantum formalism does not allow for
time being treated as an operator \cite{Jammer74}, the very status of time
in quantum theory remains a matter of debate \cite{UnruhW89}. The formalism
does not provide a precisely stated energy-time Heisenberg inequality which
should rely on a quantum commutation relation. Meanwhile, time has a
different description from space which spoils the attempts to conciliate
quantum definition of observables with relativistic behaviour under Lorentz
transformations \cite{Schrodinger30}. These inconsistencies between quantum
formalism and relativistic requirements are known to be knotty
points in the attempts to include gravity in quantum theory \cite{Rovelli91}.

Einstein introduced the principle of relativity a few months after
having proposed the hypothesis of light quanta \cite{Einstein05Quantum}.
He incidentally noticed that energy and frequency of the electromagnetic
field change in the same manner in a transformation from one inertial frame
to another \cite{Einstein05}. This remark may be considered as the first
demand for consistency between quantum and relativistic theories.
Two years later, Einstein laid down the principle of equivalence of gravity
and acceleration and predicted the existence of gravitational redshifts.
He again noticed that energy and frequency change in the
same manner under frame transformations \cite{Einstein07}.

In modern quantum theory, the similarity of energy and frequency shifts has
to be interpreted as an invariance property for particle number. This
property is well known for Lorentz transformations but usually not admitted
for transformations to accelerated frames. The latter are commonly
represented by Rindler transformations \cite{Rindler77} which do not
preserve the propagation equations of electromagnetic fields and result in a
transformation of vacuum into a thermal bath
\cite{Davies75,Unruh76,BirrellD82}. Since the
concepts of particle number and vacuum play a central role in the
interpretation of quantum field theory, this fact spoils the attempts to
interpret the Einstein equivalence principle in the quantum domain
\cite{UnruhW84,GinzburgF87}.

\section{Outline}

The basic idea underlying our approach is that relativistic effects can no
longer be described by classical relativity.
A new theoretical framework has to be built up where quantum and relativistic
requirements are treated simultaneously and consistently.
Our proposal for building up such a `quantum relativity' is to import the
conception of relativistic effects built on symmetries from classical theory
into a quantum algebraic theory.

A basic property of relativistic observables is that they undergo shifts
under frame transformations.
These shifts are however perfectly compatible with invariance properties.
In fact, they result from symmetries of the laws of physics in such a
manner that the relations between observables have their form
preserved under transformations.
To be more precise, symmetries are described by algebraic techniques which
ensure that relations between observables are universal, {\it i.e.} are
independent of the specific frame in which they are written.
Theoretical constraints associated with invariance are related to, but also
distinct from, covariance constraints corresponding to the arbitrariness
of coordinate mapping. A historical account of the relativistic approaches
emphasising respectively invariance and covariance properties may be found
in \cite{Norton93}.

\begin{figure}[htb]
\centerline{\psfig{figure=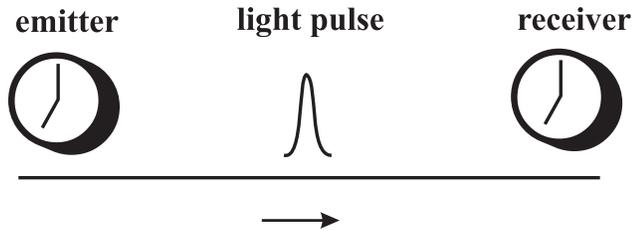,height=3cm}}
\caption{Einstein synchronisation: two remote observers, each with
a clock at his disposal, transfer a time reference by exchanging a light
pulse. }
\label{figSynchro}
\end{figure}

To make the discussion more concrete, let us consider a synchronisation
procedure where an emitter sends a light pulse to a remote receiver
(see Figure \ref{figSynchro}).
The two observers respectively encode and retrieve a time information in the
light pulse. In other words, they share a field observable which is easily
identified in a classical context as the light-cone variable $u$ defined
along the line of sight
\begin{equation}
t_e-\frac{x_e}{c}=u=t_r-\frac{x_r}{c} \label{eqSynchro}
\end{equation}
$t_e$ and $t_r$ are the emission and reception times, as delivered to
the emitter and receiver by their own clocks; $x_e$ and $x_r$ are
the space coordinates of the emitter and receiver, as measured along the
line of sight; $c$ is the velocity of light.
Clearly, such a synchronisation relies on a symmetry of physics, the
existence of a universal propagation velocity $c$. Meanwhile the time
reference shared by the two observers is a quantity preserved by field
propagation, namely the light-cone variable $u$.

\begin{figure}[htb]
\centerline{\psfig{figure=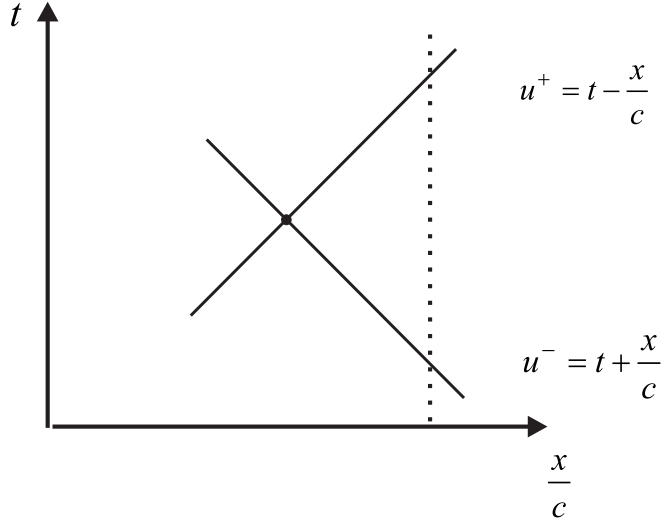,height=7cm}}
\caption{Einstein localisation represented on a space-time diagram:
space and time correspond respectively to horizontal and vertical axis;
light rays are indicated
by straight lines making a $45^{\circ }$-angle with the axis;
a motionless observer is represented by a vertical time-like trajectory;
by operating two time transfers along two different directions,
he is able to get positions in space and time of an event. }
\label{figLocal}
\end{figure}

The localisation of an event in space-time may then be operationally defined
as the result of several time transfers corresponding to different
propagation directions (see Figure \ref{figLocal}).
A motionless observer obtains the positions in time and space of an event
by measuring two light-cone variables
\begin{eqnarray}
t-\frac{x}{c}=u^{+} &\qquad &t+\frac{x}{c}=u^{-} \nonumber \\
t=\frac{u^{-}+u^{+}}2 &\qquad &\frac{x}{c}=\frac{u^{-}-u^{+}}2
\label{eqLocal}
\end{eqnarray}
It follows from the previous arguments that the relativistic notion of
space-time is ultimately based upon the Poincar\'e symmetry of field
propagation. These discussions look familiar since the invariance of
Maxwell equations under Lorentz transformations played a prime role in
Einstein's introduction of classical relativity. They have been
repeated here to prepare the reader to their quantum counterparts.

In quantum theory, the field observables can no longer be defined as classical
numbers. Taking into account this essential difference, most preceding
discussions are still relevant. In particular, the time references
used in synchronisation and localisation procedures have to be observables
preserved under propagation, that is also quantities built on the
generators of the symmetries of field propagation.
We will give below the definitions of synchronisation and localisation
observables in terms of symmetry generators.

Symmetries also play a primary role in
fundamental metrology.
Translation symmetry allows one to transport metrological standards from one
place to another.
Lorentz symmetry permits one to use standards in different inertial
frames and to derive a length unit from the time unit. The role played by
dilatation is less often discussed although the invariance of Maxwell
equations under dilatations has been known for a long time \cite{Pais82}.
Dilatations are naturally involved in comparisons of lengths or durations
with different scales.

Meanwhile the metrological definition of units is more and more evolving
towards the use of quantum standards. This is not only a result of
technological progress but, more basically, of efforts to improve the
universality of the definition of units. Dilatation symmetry plays a central
role in this context as soon as dilatation is understood as a correlated
change of time, space and mass scales which preserves the velocity of light
and the Planck constant \cite{Dicke62,Sakharov74,Hoyle75}. A correlated
variation of time and mass scales under dilatations is just the expression
of the equivalence principle or, equivalently in a metrological context, of
a consistent definition of units \cite{Guinot97}.

These ideas may be applied to accelerated frames as soon as the latter
are given a conformal representation.
The interest of this representation relies on conformal invariance
of Maxwell equations \cite{Bateman09,Cunningham09}. Moreover,
conformal coordinate transformations fit the motion of uniformly accelerated
observers like Lorentz transformations fit the motion of inertial observers
\cite{FultonRW62}.
Conformal invariance also means that propagation of electromagnetic fields is
not sensitive to a conformal variation of the metric tensor, that is a change
of
space-time scale preserving the velocity of light \cite{MashhoonG80}. Hence,
light propagates along straight lines while frequencies are preserved under
propagation. Of course, redshifts are still present since clocks rates are
affected by the conformal factor.

Conformal invariance can be rigorously established for quantum
electromagnetic fields \cite{BinegarFH83}. Moreover, the definition of
photon number is conformally invariant \cite{Gross64}. The concept of photon
and, in particular, the concept of vacuum \cite{QSO95} are thus the same for
inertial and uniformly accelerated observers which opens the way to an
extension
of `quantum relativity' to accelerated frames \cite{PRL96,PLA96,EPL97,FoP98}.

\section{Clock synchronisation}

\begin{figure}[htb]
\centerline{\psfig{figure=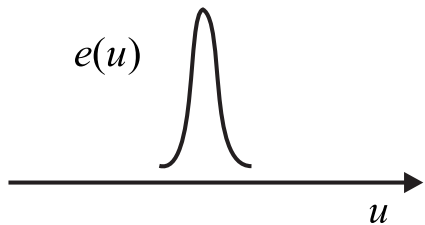,height=3cm}}
\caption{Dispersion along the line of sight of a pulse used as a time
reference in synchronisation: the energy density $e$ of the quantum field is
represented as a function of the light cone variable $u$. }
\label{figPulse}
\end{figure}

We now address the problem of clock synchronisation performed with quantum
fields. We focus our attention on quantum dispersions of energy density
along the line of sight (see eq. (\ref{figSynchro})). We may therefore
consider at this stage the simple theory of a scalar massless field
propagating along a single direction in a two-dimensional (2d) space-time.
The light pulse used as a time reference is schematically represented on
Figure \ref{figPulse}.

A free massless scalar field $\phi$ in 2d space-time is the sum of two
counterpropagating components
\begin{eqnarray}
\phi \left( x\right) &=&\varphi ^{+}\left( u^{+}\right) +\varphi ^{-}\left(
u^{-}\right) \nonumber \\
u^{+} &=&x^0-x^1 \nonumber \\
u^{-} &=&x^0+x^1
\end{eqnarray}
{}From now on, we use natural space-time units ($c=1$) and we denote $x^0$ and
$x^1$ the time and space coordinates. For the synchronisation problem,
we consider only one of the two counterpropagating components, that we
simply denote $\varphi \left( u\right)$
\begin{equation}
\varphi \left( u\right) =\int_0^{\infty }\frac{d\omega }{2\pi }\sqrt{\frac{%
\hbar }{2\omega }}\left( a_{\omega }e^{-i\omega u}+a_{\omega }^{\dagger
}e^{i\omega u}\right)
\end{equation}
$u$ is the light-cone variable and $\omega$ represents the frequency.
$a_{\omega }$ and $a_{\omega }^{\dagger }$ are the
annihilation and creation operators
\begin{equation}
\left[ a_{\omega },a_{\omega ^{\prime }}^{\dagger }\right] =2\pi \delta
\left( \omega -\omega ^{\prime }\right)
\end{equation}
$\delta$ is the Dirac distribution. The energy density $e(u)$ is defined as
a quadratic form of the fields
\begin{equation}
e(u)=:(\partial _{u}\varphi (u))^2 :
\end{equation}
The symbol $:~:$ prescribes a normal ordering of products of operators, so
that energy density vanishes in vacuum.

The total number of photons in the field state
\begin{equation}
N=\int_0^{\infty }\frac{d\omega }{2\pi }a_{\omega }^{\dagger }a_{\omega }
\end{equation}
is invariant under conformal transformations to accelerated frames.
In other words, these transformations amount to a redistribution of
particles in the frequency domain without any change of the total
number \cite{BJP95}.
Invariance of the photon number under conformal transformations
may be written
\begin{equation}
\left( E,N\right) =\left( D,N\right) =\left( C,N\right) =0 \label{invN2d}
\end{equation}
where $E$, $D$ and $C$ are the symmetry generators defined as
moments of energy density \cite{ItzyksonZ85}
\begin{eqnarray}
E &=&\int e(u)du \nonumber \\
D &=&\int ue(u)du \nonumber \\
C &=&\int u^2 e(u)du \label{defG2d}
\end{eqnarray}
$E$ is the energy-momentum, that is also the translation
operator associated with the light-cone variable $u$. $D$
corresponds to dilatations of this variable and
$C$ to transformations to accelerated frames.
Throughout the paper, quantum commutators are denoted by using the following
notation
\begin{equation}
\left( A,B\right) =\frac{1}{i\hbar }\left[ A,B\right] =\frac{AB-BA}{i\hbar }
\label{defComm}
\end{equation}
We also take care of non commutativity of operator
products by introducing a symmetrised product represented by a dot symbol
\begin{equation}
A\cdot B=\frac{AB+BA}2
\end{equation}

The commutators of the symmetry generators play a key role.
Most forthcoming computations are based on conformal algebra,
that is the set of these commutators
\begin{eqnarray}
\left( E,D\right) &=&E \nonumber \\
\left( E,C\right) &=&2D \nonumber \\
\left( D,C\right) &=&C \label{ConfAlg2d}
\end{eqnarray}

To begin with the simplest example, we write an operator $U$ defined as a
quantum analog of the classical light-cone variable $u$
\begin{equation}
U=D\cdot \frac{1}{E} \label{defU}
\end{equation}
Using the first commutation relation
in (\ref{ConfAlg2d}), one deduces the shifts of $U$ under frame
transformations associated with $E$ and $D$
\begin{eqnarray}
\left( E,U\right) &=&1 \nonumber \\
\left( D,U\right) &=&U \label{transfU}
\end{eqnarray}

These quantum shifts laws have the same form as classical expressions.
The shift under translations just means that the operator $U$ is
a time observable canonically conjugated to the energy $E$.
$U$ has the simple classical interpretation of the mean value of $u$ in
the quantum distribution $e\left( u\right)$ of Figure \ref{figPulse}.
But it has also a proper quantum definition (\ref{defU}) which holds
in any field state orthogonal to vacuum ($E\neq 0$). This definition has a
simple form because we consider a scalar field theory or, equivalently, spin-%
$0$ particles \cite{NewtonW49}.

The shifts of energy are easily derived from (\ref{ConfAlg2d})
\begin{eqnarray}
\left( E,E\right) &=&0 \nonumber \\
\left( D,E\right) &=&-E \nonumber \\
\left( C,E\right) &=&-2D=-2E\cdot U
\end{eqnarray}
Energy is preserved under translations and undergoes a shift
proportional to energy under dilatations. The shifts under dilatations
of energy and time are inverse to each other. Then, energy is shifted in a
position dependent manner under transformations to accelerated frames. This
quantum redshift law fits exactly the form of the classical Einstein law
\cite{Einstein07}. It is nevertheless written in a fully consistent quantum
framework.

At this point a few remarks are worth of consideration. The operator $U$ is
preserved under propagation like the classical variable $u$. As
explained in the introductory parts (see eq. (\ref{eqSynchro})), this is
the reason why it can be used as a time reference for transfering information
between remote observers. $U$ is preserved under propagation but shifted
under frame transformations. The laws written above express these
relativistic shifts. Meanwhile they describe also the quantum
commutation relations between observables.
This means that we have brought basic relativistic properties of
space-time observables within a quantum framework. A fact of great interest
for the physical analysis of time-frequency transfer is that these
expressions are available in the same theoretical framework where quantum
fluctuations of the various physical quantities may be analyzed. Hence, they
may be considered as setting the quantum limits in time-frequency transfer
\cite{PRL96}.

The shifts of $U$ under $E$ and $D$ as well as the transformations of $E$ under
$E$, $D$ and $C$ are identical to expectations from classical relativity.
This is no longer the case for the shift of $U$ under $C$ which may also be
derived from conformal algebra (\ref{ConfAlg2d})
\begin{eqnarray}
&&C=UEU+\frac{\alpha ^2}{E} \nonumber \\
&&\left( C,U\right) =U^2 -\frac{\alpha ^2}{E^2} \nonumber \\
&&\alpha ^2 =C\cdot E-D^2 +\frac{\hbar ^2}{4}
\end{eqnarray}
The two first relations appear as sums of a classical looking term and of a
quantum correction. The quantum corrections are written in terms of a
Casimir invariant $\alpha ^2$ of the conformal algebra \cite{PRL96}
\begin{equation}
\left( E,\alpha ^2 \right) = \left( D,\alpha ^2 \right) =
\left( C,\alpha ^2 \right) = 0
\end{equation}
This invariant has a minimum value $\frac{\hbar ^2}{4}$ which is attained
by $1$-photon states
\begin{eqnarray}
\alpha ^2 &\geq &\frac{\hbar ^2}{4} \nonumber \\
N=1 \quad \Longrightarrow \quad \alpha ^2 &=&\frac{\hbar ^2}{4}
\end{eqnarray}
As a consequence, the quantum corrections never vanish.

\section{Space-time localisation}

We come now to the problem of space-time localisation sketched on Figure \ref
{figLocal}. The basic equations (\ref{eqLocal}) mean that Einstein
localisation amounts to the transfer of two time references
along different directions. This has a simple implementation in 2d
quantum field theory since we have only to duplicate the
previous definitions for the two counterpropagating directions.
The more realistic problem of defining localisation observables
in 4d space-time will be addressed in the next section.

We introduce two sets of conformal generators, $E^{+}$, $D^{+}$ and
$C^{+}$ on one hand and $E^{-}$, $D^{-}$ and $C^{-}$ on the other hand,
which correspond to the two propagation directions.
The two sets commute with each other.
We then define two quantum light-cone variables (\ref{defU})
\begin{eqnarray}
U^{+} &=&D^{+}\cdot \frac{1}{E^{+}} \nonumber \\
U^{-} &=&D^{-}\cdot \frac{1}{E^{-}} \label{defUpm}
\end{eqnarray}
and interpret them as defining a position in time and a
position in space as in classical equations (\ref{eqLocal})
\begin{eqnarray}
X^0 &=&\frac{U^{-}+U^{+}}2 \nonumber \\
X^1 &=&\frac{U^{-}-U^{+}}2 \label{defX01}
\end{eqnarray}
Clearly, these observables are associated with the physical event
defined by the intersection of the two light pulses of Figure \ref{figLocal}.
These observables obey canonical conjugation relations with momentum operators.
More generally, the shifts they undergo under the action of $E^{+}$, $E^{-}$,
$D^{+}$ and $D^{-}$ have a classical form.

To the aim of rewriting these results in an explicitly Lorentz covariant
manner, we introduce generators $P_\mu$ which represent translations along
the various axis and, also, generators $J_{\mu \nu}$ for rotations, $D$
for dilatation and $C_\mu$ for conformal transformations to uniformly
accelerated frames. Equations (\ref{defUpm}) are thus rewritten
\begin{eqnarray}
J^{\mu \nu} &=&P^\mu \cdot X^\nu -P^\nu \cdot X^\mu \nonumber \\
D &=&P_\mu \cdot X^\mu \equiv P\cdot X \label{JDclass}
\end{eqnarray}
Meanwhile position observables (\ref{defX01}) are read
\begin{equation}
X^\mu =\frac{P^\mu }{P^2}\cdot D-\frac{P_\nu }{P^2}\cdot J^{\mu
\nu }
\label{defX}
\end{equation}
The Minkowski tensor $\eta ^{\mu \nu}$ is used to raise or lower indices
with a $+$ signature for time components and a $-$ signature for space ones.

In (\ref{defX}), $P^2$ is the squared mass associated with the field
state according to the usual relativistic definition \cite{Einstein06}
\begin{equation}
P^2 =P_\nu P^\nu
\end{equation}
$P^2 =E^{+} E^{-}$ differs from zero as soon as the field contains energy
propagating in the two different propagation directions.
Space-time positions $X^\mu$ may be defined in this case only.
A vanishing mass indeed indicates a field state with a single propagation
direction which can be used for synchronisation but not for localisation
purposes.

The definition (\ref{defX}) of space-time positions associated with the
field state is quite analogous to Einstein's classical definition of spatial
positions \cite{Einstein06}. However, it involves not only the rotation
generators $J^{\mu \nu}$ but also the dilatation generator $D$. As a
result, a position in time $X^0$ is defined together with a position in
space $X^1$. Furthermore, these definitions hold in the quantum domain,
with the particularly important outcome that the space-time observables are
canonically conjugated to energy-momentum operators. We will come back to
these properties after having given a more general treatment of Einstein
localisation.

\section{Localisation and spin}

The description of Einstein localisation given in the previous section
heavily relies on a specific feature of 2d field theories, namely the
existence of an {\it a priori} decomposition of fields in counterpropagating
directions. In 4d space-time in contrast, such a natural decomposition is
not available. Furthermore, light rays have an intrinsic transverse
extension due to diffraction and two light rays do not necessarily cross
each other. The description of localisation procedures may nonetheless be
given following the same line of thought.

Poincar\'e transformations are now described by $10$ generators, namely the
$4$ components $P_\mu$ representing translations and the $6$ independent
components of the antisymmetric tensor $J_{\mu \nu}= -J_{\nu \mu }$
representing
rotations and Lorentz boosts. The commutators between
these symmetry generators constitute the Poincar\'e algebra
\begin{eqnarray}
&&\left( P_\mu ,P_\nu \right) =0 \nonumber \\
&&\left( J_{\mu \nu},P_\rho \right) =\eta _{\nu \rho }P_\mu -\eta _{\mu
\rho }P_\nu \nonumber \\
&&\left( J_{\mu \nu},J_{\rho \sigma }\right) =\eta _{\nu \rho }J_{\mu
\sigma }+\eta _{\mu \sigma }J_{\nu \rho }-\eta _{\mu \rho }J_{\nu \sigma
}-\eta _{\nu \sigma }J_{\mu \rho } \label{PAlg}
\end{eqnarray}
This algebra has two Casimir invariants, the squared mass $P^2$
and the squared spin $S^2$. Spin observables are introduced in a Lorentz
covariant manner through the Pauli-Lubanski vector $W^\mu$
\begin{eqnarray}
W^\mu &\equiv &-\frac{1}2 \epsilon ^{\mu \nu \rho \sigma }J_{\nu \rho
}P_{\sigma } \nonumber \\
\left( P_\mu ,W_\rho \right) &=&0 \nonumber \\
\left( J_{\mu \nu},W_\rho \right) &=&\eta _{\nu \rho }W_\mu -\eta _{\mu
\rho }W_\nu \label{PW}
\end{eqnarray}
$\epsilon _{\mu \nu \lambda \rho }$ is the completely antisymmetric Lorentz
tensor \cite{ItzyksonZ85}
\begin{eqnarray}
\epsilon _{{\rm 0123}} &=&-\epsilon ^{{\rm 0123}}=+1 \nonumber \\
\epsilon _{\mu \nu \rho \sigma } &=&-\epsilon _{\mu \nu \sigma \rho
}=-\epsilon _{\mu \rho \nu \sigma }=-\epsilon _{\nu \mu \rho \sigma }
\end{eqnarray}
The commutators between components of the spin vector may be written in
terms of a spin tensor $S_{\mu \nu}$
\begin{equation}
\left( W_\mu ,W_\nu \right) =P^2 S_{\mu \nu}=\epsilon _{\mu \nu \rho
\sigma }W^\rho P^{\sigma }
\end{equation}
The spin tensor can be extracted from this equation only for a non vanishing
mass. Spin observables commute with momentum and they are transverse with
respect to momentum
\begin{equation}
P^\mu S_{\mu \nu}=P_\mu W^\mu =0
\end{equation}
The squared spin is a Lorentz scalar that we can write in
its standard form in terms of a spin number $s$ taking integer or
half-integer values
\begin{equation}
S^2 =\frac{W^2}{P^2}=\frac{1}2 S_{\mu \nu}S^{\nu \mu }=-\hbar
^2 s\left( s+1\right)
\end{equation}

Dilatation symmetry is described by enlarging Poincar\'e algebra
(\ref{PAlg}) by a generator $D$
\begin{equation}
\left( D,P_\mu \right) =P_\mu \qquad \left( D,J_{\mu \nu}\right) =0
\label{DAlg}
\end{equation}
Generally speaking, commutation relations with $D$ define
the conformal weight of observables. This weight vanishes for $%
J_{\mu \nu}$ but not for $P_\mu$.

To build up position observables, we write quantum generalisations
of equations (\ref{JDclass})
\begin{eqnarray}
J^{\mu \nu} &=&P^\mu \cdot X^\nu -P^\nu \cdot X^\mu + S_{\mu \nu}
\nonumber \\
D &=&P_\mu \cdot X^\mu \equiv P\cdot X
\label{JD}
\end{eqnarray}
The angular momenta $J_{\mu \nu}$ are now sums of orbital and spin
contributions. They fix the part of position observables transverse to
momentum while the expression of $D$ determines their longitudinal part.
As soon as the field contains photons propagating in at least two
different directions, the squared mass differs from zero and
equations (\ref{JD}) may be solved to obtain the space-time
observables. Their expression remains identical to (\ref{defX}).

The shifts of these observables under translations, dilatation and
rotations are shown from (\ref{PAlg},\ref{DAlg}) to fit
exactly the shifts of coordinate parameters under the
corresponding transformations of classical relativity
\begin{eqnarray}
\left( P_\mu ,X_\nu \right) &=&-\eta _{\mu \nu} \nonumber \\
\left( D,X_\mu \right) &=& -X_\mu \nonumber \\
\left( J_{\mu \nu},X_\rho \right) &=&\eta _{\nu \rho }X_\mu -\eta _{\mu
\rho }X_\nu \label{PX}
\end{eqnarray}
The first equation also means that observables $X_\mu$ are conjugated to
energy-momentum operators. This entails that canonical commutation relations
are embodied in the symmetry algebra.

Commutators between different components of positions (\ref{defX}) may also
be deduced
\begin{equation}
P^2 \cdot \left( X_\mu ,X_\nu \right) =S_{\mu \nu}
\end{equation}
These commutators do not vanish in general which
is reminiscent in the present approach of the known
problem of localisability of particles with spin \cite{Pryce48,Fleming65}.
Clearly concepts originating from classical conceptions of space-time have
to be modified in a fully quantum theoretical framework.

The observable $X_\mu$ is a position in time for $\mu =0$ and
in space for $\mu =1,2,3$. All definitions and relations written above obey an
explicit Lorentz covariance. In particular a time observable has been
defined which is conjugate to energy in the same manner as space observables
are conjugate to spatial momenta. Observables are built on conserved
quantities and do not evolve due to field propagation. In
particular, the time observable $X_0$ represents a date, that is
the position of an event in time.

Once again, position observables can be defined only
when the squared mass does not vanish. Therefore, the domain of
definition of localisation observables does not cover the space of all field
states so that these hermitic observables are not self-adjoint \cite
{BogolubovLT}. This does not forbid one to build up a rigorously consistent
treatment as exemplified by the formalism of positive operator valued
measures \cite{LevyLeblond76,BuschGL95}. The present paper
is based on a quantum algebraic calculus operating in the algebra of
observables. This calculus is rigorously defined as soon as
divisions by $P^2$ are manipulated with care which, of course, restricts
the domain of validity of some relations to massive states \cite{Toller98}.

\section{Redshifts and metric factors}

As already discussed, conformal symmetry allows us
to deal with accelerated frames. To this aim, we introduce
$4$ additional conformal generators $C_\mu$
which represent transformations to accelerated frames.

Conformal algebra contains commutators (\ref{PAlg},\ref{DAlg})
complemented by the following ones
\begin{eqnarray}
&&\left( C_\mu ,C_\nu \right) =0 \nonumber \\
&&\left( D,C_\mu \right) =-C_\mu \nonumber \\
&&\left( P_\mu ,C_\nu \right) =-2\eta _{\mu \nu}D-2J_{\mu \nu}
\nonumber \\
&&\left( J_{\mu \nu},C_\rho \right) =\eta _{\nu \rho }C_\mu -\eta _{\mu
\rho }C_\nu \label{CAlg}
\end{eqnarray}
The generators $C_\mu$ are commuting components of a Lorentz vector
with a conformal weight opposite to that of momenta.
Commutators $\left( P_\mu ,C_\nu \right)$ describe the redshifts of
momenta and thus constitute quantum versions of the Einstein redshift law.

We also introduce the generic generator $\Delta_a$ of transformations to
accelerated frames
\begin{equation}
\Delta_a =\frac{a \cdot C}2
\end{equation}
where the vector $a^\mu$ contains accelerations along four
space-time directions. The redshift of squared mass has exactly
the form expected from Einstein classical law \cite{EPL97}
\begin{equation}
\left( \Delta_a ,P^2 \right) =2 P^2 \cdot \left( a \cdot X \right)
\label{CM}
\end{equation}
It is indeed proportional to $P^2$ and to a gravitational potential
$a \cdot X$ depending linearly on the position measured along the
acceleration. It may also be read as a conformal metric factor arising in
transformations to accelerated frames and depending on observables
$X$ as the classical metric factor depends on classical coordinates
\cite{FoP98}.

In contrast the redshifts of momenta differ from the
classical law since they depend on spin observables
\begin{eqnarray}
\left( \Delta_a ,P_\nu \right) &=& a_\nu D - a^\mu J_{\mu \nu}
\nonumber \\
&=&a_\nu P\cdot X-a^\mu P_\mu \cdot
X_\nu +a^\mu X_\mu \cdot P_\nu-a^\mu S_{\mu \nu} \label{CP}
\end{eqnarray}
When applied to momenta, Einstein redshift law
should therefore be regarded as a classical
approximation valid in the limiting case where
spin contributions are negligible.
Notice that spin dependence disappears in the mass redshift (\ref{CM})
as a consequence of transversality of spin and momentum vectors.
Both redshift laws (\ref{CM}-\ref{CP}) have a universal form dictated by
conformal algebra, although the latter form differs from the classical one.

Interesting insights on the universality of relativistic transformations
are obtained as consequences of the preservation of canonical commutators
under frame transformations.
Precisely the commutators $\left( P_\mu ,X_\nu \right)$ are classical
numbers which commute with generators $\Delta_a$
\begin{eqnarray}
&&\left( \Delta_a , \left( P_\mu ,X_\nu \right) \right) =0
\end{eqnarray}
The following identity is then obtained from the Jacobi identity,
a general consequence of (\ref{defComm})
\begin{eqnarray}
&&\left( \left( \Delta_a ,P_\mu \right) ,X_\nu \right) =
\left( \left( \Delta_a ,X_\nu \right) ,P_\mu \right)
\label{Identity1}
\end{eqnarray}
We may also evaluate the first expression from (\ref{CP})
\begin{equation}
\left( \left( \Delta_a ,P_\mu \right) ,X_\nu \right) =
-a_\mu X_\nu -\eta _{\mu \nu} a \cdot X + a_\nu X_\mu
\label{Identity2}
\end{equation}
Identity (\ref{Identity1}) proves that the relativistic transformations of
space-time
scales and energy-momentum redshifts are consistent with each other.
It thus extends to quantum relativity a set of consistency rules which
are well known in classical relativity. Furthermore identity (\ref{Identity2})
shows that these expressions have a classical form although the
shifts $\left( \Delta_a, P_\mu \right)$ and $\left( \Delta_a, X_\nu \right)$
both differ
from classical predictions.

An even more remarkable result is obtained when the previous equations
are symmetrised in the exchange of the two indices $\mu$ and $\nu$
\begin{eqnarray}
\left( \left( \Delta_a ,P_\mu \right) ,X_\nu \right) +\left( \left(
\Delta_a ,P_\nu \right) ,X_\mu \right) &=&-2\eta ^{\mu \nu}a\cdot X
\nonumber \\
\left( \left( \Delta_a ,X_\nu \right) ,P_\mu \right) +\left( \left(
\Delta_a ,X_\mu \right) ,P_\nu \right) &=&-2\eta ^{\mu \nu}a\cdot X
\end{eqnarray}
The last relation has exactly the form of the classical definition of the
metric factor. As a matter of fact, $\left( \Delta_a ,X_\mu \right)$ is
the shift of position under the generator $ \Delta_a$ and $\left( \left(
\Delta_a ,X_\mu \right) ,P_\nu \right)$ is the variation of this
shift under an infinitesimal translation. The resulting expression only depends
on
the conformal factor $a\cdot X$ which already appears in the
mass redshift (\ref{CM}). This factor is identical to the classical expression
but now written in terms of quantum positions.

In the particular case $\mu =\nu =0$ for example, the preceding
expressions give informations about the redshifts of energy and of
clock rates. Yet these properties have been derived
from conformal symmetry without the addition of
any further assumption, like the `clock hypothesis' of classical
relativity. This means that conformal symmetry is sufficient to force
properly defined observables to have their relativistic transformations
determined by the metric factors of classical relativity.

Let us now summarise the main results which have been obtained
in this new `quantum relativity' framework.
An algebra of quantum observables has been defined as the enveloping
division ring built upon the symmetry algebra.
Quantum and relativistic properties have then been obtained through
algebraic computations which naturally embody symmetry properties.
In particular, this quantum algebraic calculations have allowed us to
define the localisation observables, write down their commutation relations,
derive their relativistic shifts and to begin to describe
metric properties in a quantum theoretical framework.

\end{document}